\documentclass{evn2004}
\usepackage{txfonts}
\usepackage{natbib}
\usepackage{graphicx}
\begin{document}
   \title{Signatures of restarted activity in core-dominated triples}

   \author{Andrzej Marecki\inst{1}}

   \institute{Toru\'n Centre for Astronomy, N. Copernicus University,
           PL-87-100 Toru\'n, Poland}

   \abstract{The lobes of radio-loud AGNs are huge reservoirs of energy so
even if the central engine turns off, the radio source is still observable
for a substantial amount of time. It is quite obvious, however, that the
ignition of a new period of activity can occur before the lobes have faded
completely. Restarted activity has been observed in several giant radio
galaxies but until recently, it has hardly ever been observed in more
compact sources. Here we present an early result of a pilot VLBA programme
targeted at compact core-dominated triples. It is found that the
milliarcsecond structure of the central components in that type of objects
normally take the form of core-jets, which in many cases do {\em not}
point toward the relic lobes. Therefore, it looks as if the phenomenon of
the restarted activity can occur regardless of the source size and it is
often accompanied by a repositioning of the central engine. }

   \maketitle
%

\section{Introduction}

The large linear sizes of the majority of radio sources, which, for some radio
galaxies termed Giant Radio Galaxies (GRGs), can be more than 1\,Mpc, are a
consequence of the constant growth of the radio structures during an
appreciable period of time. According to \citet{al87} and \citet{liu92}
the spectral ages of extended radio sources are $\sim$$10^7$ to
$10^8$~years. However, even in the case of a GRG the spectral age is two
orders of magnitude less than the age of the galaxy itself which, in
general, can be of the same order as the age of the Universe; the Milky
Way Galaxy is a good example of this \citep{pasq04}. It seems that the
phenomena collectively called `activity' are episodic --- galaxy evolution
can be interrupted at some stage and then it can continue after a certain
period of time.

If the energy supply from the central engine to the hotspots and the lobes
cuts off, the radio source is still observable for a substantial period of
time: $10^8$~yr \citep{kg94}. Therefore, it is quite likely that a new
epoch of activity can occur before the lobes have faded completely. The
observable effect of this would be the presence of a new, bright
component(s) located in or straddling the centre of a larger, double-lobed
relic structure. The signature of the renewed activity is most obvious if
it takes the form of a smaller, double-lobed radio source giving rise to
the so-called double--double structure \citep{schoen00a,schoen00b}.
J0116$-$473 \citep{sarip02} and PKS\,B1545$-$321 \citep{sarip03} are also
clear examples of double--double radio galaxies (DDRG).

Alternatively, `restarted activity' may result in `X-shaped' radio sources
\citep[and references therein]{helgephd} such as 3C\,223.1 or 3C\,403
\citep{dt02}. It can be argued that these two sources resemble DDRGs such
as J0116$-$473 or PKS\,B1545$-$321 except for a misalignment between the
inner (active) and the outer (inactive) parts. The mechanism that
triggered the development of the new structure in these two kinds of
objects could be the same but that DDRGs are just `special cases' in which
the misalignment is (close to) zero.

In principle, different mechanisms could be responsible for activity
renewal in particular types of objects, especially if they depend on the
presence/absence of the misalignment between the old and the new
structures and the magnitude of the ratio of those sizes. However, mergers
seem to be the most `natural' explanation of the activity. A review of
other possible mechanisms has been given by \citet{schoen00a}.

All the objects in the examples quoted above are large with linear sizes
of the order of several hundreds kpc. It would appear that radio sources
have to be quite old for activity renewal regardless of their
morphological features. The apparent lack of observable double--double or
X-shaped structures in small-scale (i.e. young) sources could mean that
there are inherent limits in the mechanism(s) of activity re-ignition
known to exist in GRGs. This could make it rare if not impossible for such
phenomenon to take place in compact objects. Alternatively, the physical
conditions inside the cocoon might not favour the development of the inner
lobes for a long time after the initial burst of the activity. According
to \citet{kaiser00} that timescale could be up to $\sim$$5 \cdot 10^7$
years so, even if the activity in a source younger than that actually
ceased and then restarted, such events would remain unobservable.

\section{A search for compact restarted RLAGNs}

So, we are facing an interesting question: do RLAGNs whose activity stops
and is restarted on timescales as low as e.g. $10^5$\,years exist at all?
If so, they should be compact and have a clear signature of restarted
activity, i.e. a bright core and weak, diffuse lobes without hotspots. The
first example of an object of this kind --- 0402+379 --- was given by
\citet{maness03}. On arcsecond scales this source is a core-dominated
Medium-scale Symmetric Object (MSO) with very steep spectrum lobes. The
multi\-frequency VLBA observations carried out by \citet{maness03} provide
impressive evidence of intermittent activity: the innermost double ($\sim$
30\,pc across) has a FR\,II type structure. However, some hints of a
previous active epoch are still preserved. In particular, the diffuse
component $\sim$200\,pc south of the centre seems to be a remnant of the
past active period. Altogether there are three pairs of lobes visible in
this source, two `dead' and one `alive'. The authors suggest that the
intermittency in 0402+379 is caused by the presence of two orbiting black
holes, although they find inconsistencies in that model.

The case of 0402+379 is of particular importance. It is only the third
example of a Compact Symmetric Object (CSO) possessing a large-scale
structure but the first where ``large'' scale is actually only a
kiloparsec scale. The other two are 0108+388 \citep{baum90} and 1245+676
\citep{mar03}. Finding more of these objects would be essential if more
light is to be shed on the mechanisms of AGN ignition. Therefore, we
started a search programme to find similar sources or at least good
candidates for frequently restarting MSOs based on publicly available
data. Once candidates had been identified, their detailed structures could 
be determined using MERLIN and/or VLBI.

There are several possible ways to find candidate sources. What we did was
to inspect the maps of `rejects' i.e. sources not suitable to be used as
cali\-brators as they did not appear pointlike in the Jodrell Bank--VLA
Astrometric Survey (JVAS) \citep{jvas1, jvas2, jvas3}. We found 13 objects
(Table~\ref{t-jvas-cand}) in which there was a dominant core straddled by
(nearly) symmetric outer components which are the putative lobes pertinent
to a previous stage of activity.

\begin{table}[h]
\caption[]{Candidates for restarted RLAGNs among JVAS rejects}
\begin{tabular}{c c c} \hline

J0458+2011 & J1708+0035 & J1848+3244\\
J1012+3309 & J1756+2914 & J1857+3104\\
J1520+5635 & J1803+0934 & J2005+1825\\
J1628+2247 & J1843+3225 & J2218+4146\\
J1647+2705 & & \\

\hline
\end{tabular}
\label{t-jvas-cand}
\end{table}

There were no published VLBI maps of the sources listed in
Table~\ref{t-jvas-cand} but as JVAS provided a basis for the VLBA
Calibrator Survey \citep[VCS,][]{vlbacs} a check was made using the VCS
server\footnote{http://magnolia.vlba.nrao.edu/vlba\_calib.} to see if any
of the sources, although being JVAS rejects, had been mapped with the VLBA
and the resulting images included in the VCS. As expected, quite a few had
been included and one of them --- J1708+0035 --- seemed to be particularly
interesting.

\section{The case of J1708+0035}

\begin{figure}[ht]
\centering
\includegraphics[scale=0.5]{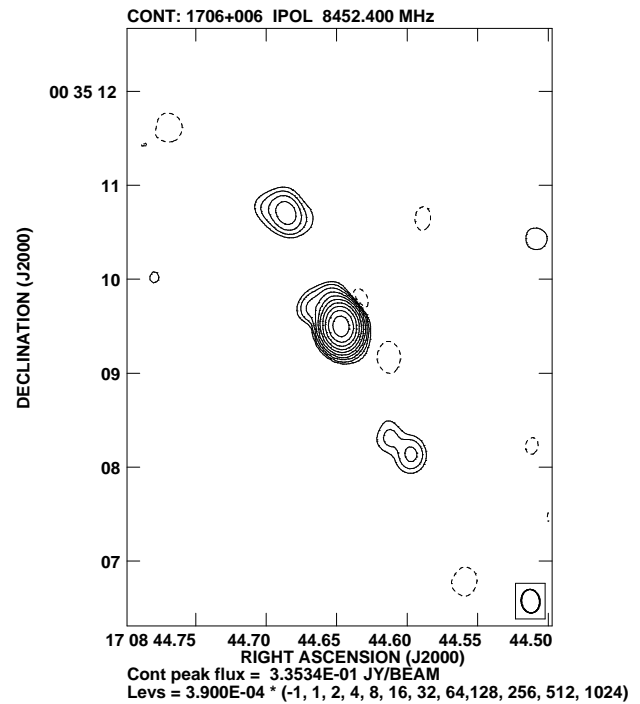}
\caption{JVAS map of J1708+0035 acquired with the VLA in A-configuration
at 8.4\,GHz.}
\label{JVAS_map}
\end{figure}

J1708+0035 is a galaxy located at a redshift of $z=0.449$. In the JVAS map
\citep{jvas2} it appears as a core-dominated aligned triple at
P.A.$\approx25\degr$ (Fig.~\ref{JVAS_map}). Its projected linear size is
$\sim$10\,$h^{-1}$\,kpc.

\begin{figure}[t]
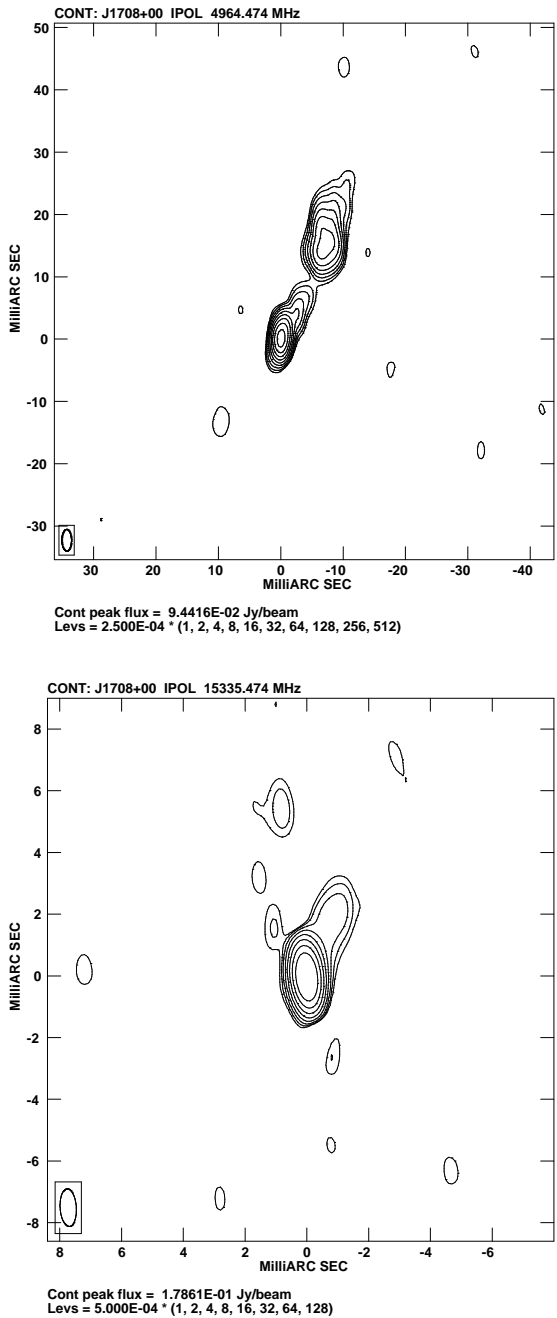

\centering
\includegraphics[scale=0.4]{J1708+0035.ICL001.6CM.PS}
\vspace{1 cm}
\includegraphics[scale=0.4]{J1708+0035.ICL001.2CM.PS}
\caption{VLBA maps of J1708+0035 at 5\,GHz (top) and 15\,GHz (bottom).
Note the different scales on both maps. The north-western feature seen in 
the 5\,GHz image does {\em not} show up on the 15\,GHz image at all so its
absence in the map shown does not result from the enlargement of the
central core-jet feature.}
\label{VLBA_maps}
\end{figure}

The quality of the VCS maps is rather poor and it is unclear whether this
particular source is either a double or a core-jet. Nevertheless, whatever
the structure actually is, there is a hint in the VCS maps that it might
be {\em misaligned} with respect to that seen in the JVAS map
\citep{jvas2}. Therefore, we observed J1708+0035 with the VLBA at two
frequencies (5\,GHz and 15\,GHz) and, using standard reduction techniques
with phase referencing followed by self-calibration, produced the maps
shown in Fig.~\ref{VLBA_maps}. At 15\,GHz the source is a core-jet whereas
at 5\,GHz an additional steep-spectrum part of the jet is also visible.
The whole milliarcsecond structure points to P.A.$\approx-25\degr$
confirming the existence of a large misalignment between the arcsecond and
milliarcsecond structures, tentatively suggested by the VCS maps.

\section{Discussion}

The phenomenon of misalignments between milli\-arcsecond- and
arcsecond-scale structures in radio sources is well known and has, in
fact, been thoroughly investigated --- see \citet{appl96} for a review.
Based on a large collection of observational material, they have concluded
that, although it is possible to apply a helical jet model to individual
sources, it is difficult to find a single mechanism responsible for the
observed distribution of misalignments indicated by a statistically
significant peak at $\Delta PA\simeq 90\degr$. This peak is particularly
well defined for BL~Lac objects. One way to interpret the misalignments is
to adopt a complicated twisted helical jets model. This has been done for
example by \citet{arw99} for the quasar TXS\,1055+018 where the jet is
pointing exactly perpendicular to the whole structure of the radio
source. More recently, \citet{cass02} carried out VLBI observations of
several BL~Lac objects in 1-Jy Sample, in order to investigate the
spatial evolution of radio jets from a few tens to hundreds of
milliarcseconds, and to search for helical jets in this class of sources.
Their EVN and MERLIN observations of bent radio jets in BL~Lac objects do
not lend firm support to the explanation of misalignments by means of
helical jets.

Thus, although an interpretation of misaligned sources based on a twisted
jets concept can be argued, adopting a scenario of re-ignition with a
simple change of jet axis leads to a straightforward and natural
explanation without complex modelling. Re-ignition becomes even more
plausible when the overall linear size of the source in question is small.
In the case of TXS\,1055+018, which has a size of 150\,kpc\,$h^{-1}$, a
helical model might be viable. However, in the case of J1708+0035 which is
15 times more compact there is simply not enough space for the helix to be
squeezed into it.

Finally, we would like to suggest that misaligned sources could actually
be compact X-shaped sources in which one `arm' of the cross lies almost in
the sky plane and the other `arm', namely the one pertinent to the reborn
radio source, is highly beamed toward the observer. But do compact
X-shaped sources exist? In fact, such a class has not been recognised so
far. Nevertheless, we {\em have} serendipitously found a map of a compact
X-shaped source (TXS\,0229+132) in the literature \citep{murphy93}.
According to Browne (priv. comm.) it is very unlikely that the strange
structure of this source could result from gravitational lensing. It is,
however, easily imaginable that if one arm of its cross-like structure was
pointed towards the observer, the image ``distorted'' by Doppler boosting
would show the pattern we have here: a core-dominated triple. We claim
that, taking into account the difficulties of the helical jet model when
trying to explain large misalignments \citep[and references
therein]{appl96,cass02}, the restarted activity concept is a very
competitive alternative because of its simplicity and plausibility.

\begin{acknowledgements}

I thank Ian Browne for sharing the JVAS data for J1708+0035 from which we
produced the image shown in Fig.~\ref{JVAS_map}, and Peter Thomasson for
careful reading of the manuscript.

\end{acknowledgements}

\end{document}